\documentclass[12pt,epsf]{article}
\usepackage{graphicx, amsmath}

\begin{document}
\sloppy
\begin{flushright}{SIT-HEP/TM-55}
\end{flushright}
\vskip 1.5 truecm
\centerline{\large{\bf Remote Inflation: }}
\centerline{\large{\bf Hybrid-like inflation without
hybrid-type potential}}
\vskip .75 truecm
\centerline{\bf Tomohiro Matsuda\footnote{matsuda@sit.ac.jp}}
\vskip .4 truecm
\centerline {\it Laboratory of Physics, Saitama Institute of Technology,}
\centerline {\it Fusaiji, Okabe-machi, Saitama 369-0293, 
Japan}
\vskip 1. truecm

\makeatletter
\@addtoreset{equation}{section}
\def\theequation{\thesection.\arabic{equation}}
\makeatother
\vskip 1. truecm

\begin{abstract}
\hspace*{\parindent}
A new scenario of hybrid-like inflation is considered without using
 hybrid-type potential. 
Radiation raised continuously by a dissipating inflaton field keeps
 symmetry restoration in a remote sector, and the false-vacuum energy of
 the remote sector dominates the energy density during inflation. 
Remote inflation is terminated when the temperature reaches the critical
 temperature, or when the slow-roll condition is violated. 
Without introducing a complex form of couplings, inflaton field may
 either roll-in (like a standard hybrid inflation) or roll-out 
 (like an inverted-hybrid model or quintessential inflation) on arbitrary
 inflaton potential. 
Significant signatures of remote inflation can be observed in the
 spectrum caused by (1) the inhomogeneous phase transition in the remote
 sector, or (2) a successive phase transition in the remote sector. 
Remote inflation can predict strong amplification or suppression of
 small-scale perturbations without introducing multiple inflation. 
Since the inflaton may have a run-away potential, it is also possible to
 identify the inflaton with quintessence, without introducing additional
 mechanisms. 
Even if the false-vacuum energy is not dominated by the remote sector,
 the phase transition in the remote sector is possible during warm
 inflation, which may cause significant amplification/suppression of the
 curvature perturbations.   
\end{abstract}

\newpage
\section{Introduction}
From recent cosmological observations, inflation appears to be the most
successful cosmological model giving the large scale structure of the
Universe that existed in the very early Universe. 
Current observations of the temperature anisotropy of the cosmic
microwave background (CMB) suggest that an inflationary Universe is
consistent with the present Universe. 
Although it is not always necessary, a typical slow-roll
condition is an important ingredient for the inflationary
scenario. 
Considering slow-roll inflation, there are two possible
dynamics scenarios for velocity damping: the original (supercooled)
inflation scenario in which the friction is induced by the Hubble
parameter, and the so-called warm inflation scenario in which a strong
diffusion produces significant friction for the inflaton
motion\cite{warm-inflation-original}.  
In the strongly dissipative warm inflation (SDWI) scenario, dissipative
 effects dominate the friction term.
If the dissipation is not so strong as to dominate the friction term 
while the radiation production occurs concurrently during the inflationary
 period\cite{Hosoya-Sakagami}, this scenario is termed the weakly
 dissipative warm inflation (WDWI) scenario.
The conventional criteria for discriminating warm inflation from
supercooled inflation is not $\Gamma > 3H$ in the friction term
 but $T> H$ for the temperature during inflation, where $\Gamma, H, T$
 denote the damping rate, Hubble parameter and temperature.
In this paper we consider the significant thermal effect of the
 radiation created during warm inflation, for both SDWI ($\Gamma > 3H$)
and WDWI ($\Gamma < 3H$) scenarios. 
Two different sources are considered for the radiation:
one is supplied continuously from the inflaton field, and the other is
supplied temporarily from the phase transition and the decay in the
remote sector.

Originally, the spectrum of the cosmological perturbations created during
inflation has been expected to be scale-invariant and Gaussian.
However, recent observations may suggest that small anomalies are
significant in the spectrum, such as a small departure from exact
scale-invariance and a certain non-Gaussian character
\cite{Bartolo-text, NG-obs}, both of which can help 
reveal the underlying theory based on the expected form of the
inflationary action.
In particular, observation of a tiny shift in the spectrum
index $n-1\ne 0$ is an obvious example\cite{EU-book} of
scale invariance violation that can be used to determine the strict
form of the inflaton potential for the inflation scenario.
Observation of non-Gaussian character in the spectrum may lead to a
significant bound for the mechanism of generating curvature
perturbations.
Determining how these anomalies arise and how they are related to
the inflation dynamics and also to the generating mechanism of the
cosmological perturbations will depend on the cosmological model and the
character of 
the underlying theory. 
In order to explain how these anomalies are explained, there are
already many models of inflation and the generating mechanism of the 
curvature perturbations, in
which the spectrum is generated 
(1) during inflation \cite{Modulated-matsuda, Kofman-modulated, A-NEW,
matsuda-stop-index, roulette-inflation},
(2) at the end of 
inflation \cite{End-Modulated, End-multi, End-multi-mat0, End-multi-mat},
  (3) after inflation by preheating\cite{IH-PR, kin-NG-matsuda, 
IH-string} and reheating\cite{IH-R, Preheat-ng}    
 or (4) late after inflation by curvatons \cite{curvaton-paper,
 curvaton-dynamics, matsuda_curvaton}
or by inhomogeneous phase transition\cite{IH-pt}. 
The idea of inhomogeneous phase transition is very useful and unifies
inhomogeneous scenarios, as it
explains density perturbations from a spatially inhomogeneous transition
accompanied by a gap in the scaling relation; namely 
$\rho\propto a^{-A}\rightarrow \rho\propto
a^{-B}$ with $A\ne B$, where $a$ denotes the cosmic-scale factor.
For example, $(A,B)=(3,4)$ and $(A,B)=(0,4)$ reproduces inhomogeneous
reheating\cite{IH-R} and creation of perturbations at the end of
inflation\cite{End-multi}, respectively. 
In this paper we consider (1) and (2),
 namely the creation of the curvature
perturbations during inflation and at the end of inflation.
However, contrary to the conventional model, the inhomogeneities at the
end of inflation are not induced by the fluctuations of the 
end-boundary of the inflaton field ($\delta \phi_e\ne 0$) but by the 
inhomogeneous phase transition in the remote sector based on
spatial inhomogeneities of the critical temperature ($\delta T_c\ne 0$).

\subsection{Cosmological perturbations in warm inflation}
Recently, the origin of the cosmological perturbations in warm inflation
has been revealed in Ref.\cite{matsuda-warm}.
In this paper, we follow the discussions in Ref.\cite{matsuda-warm}.
In warm inflation, the effective potential of
the inflaton fields $\phi_i$ depends on temperature $T$, which can be
expressed as $V(\phi_i,T)$. 
The damping rate $\Gamma_i$ of the inflaton fields
may also depend on $T$ and $\phi_i$. 
Therefore, the ``trajectory'' of warm inflation
is generally given by inflaton fields $\phi_i$ and $T$.
There may be multiple trajectories for inflation, and the correct one
cannot be chosen independent of the initial condition that is generally
given by $\phi_i$ and $T$. 
The end-boundary of inflation may also be determined by $\phi$ and $T$.
The situation is similar to multi-field models of inflation.

To compare the situations for multi-field inflation and warm
inflation, first consider two-field inflation with a  
fast-rolling inflaton $\phi_F$ that reaches its minimum during
inflation.
We assume that the slow-rolling inflaton $\phi$ couples to the trigger
field in the hybrid-type potential and this field determines the
end of inflation.
In this case, the fluctuation of the e-foldings number 
$\delta N\simeq H\delta t$ is expected to be given by 
$\delta \phi$ and $\dot{\phi}$, which naturally leads to\cite{End-multi-mat0} 
\begin{equation}
\label{bai-ve}
\delta N \sim H\frac{\delta \phi}{\dot{\phi}}.
\end{equation}
However, for  multi-field inflation the usual definition of
the curvature perturbations on spatially flat hypersurfaces
is given by using $\rho$(density),
$P$(pressure) and $\delta q$(momentum perturbation).
Following the standard definition of the curvature perturbations, the
curvature perturbations are found to be given by
\begin{eqnarray}
{\cal R}_{multi}^{(ini)}=-H\frac{\delta q}{\rho+P}=
H\frac{\dot{\phi}_F\delta \phi_F+\dot{\phi}_s \delta \phi_s}
{\dot{\phi}_F^2+\dot{\phi}_s^2}
\simeq H\frac{\dot{\phi}_s \delta \phi_s+\dot{\phi}_F \delta \phi_F}
{\dot{\phi}_F^2},
\end{eqnarray}
where $\dot{\phi}_F\gg \dot{\phi}$
on the steep potential is assumed.
In this case, the evolution during inflation ($\Delta {\cal R}$)
is crucial.
Namely, at a ``bend'' in the inflation trajectory, where the
fast-rolling field $\phi_F$ reaches its minimum, the correction caused
by the multi-field evolution is given by 
\begin{eqnarray}
\left(\Delta{\cal R}_{multi}\right)^2\simeq 
\left(H\frac{\delta \phi}{\dot{\phi}}\right)^2
\gg \left({\cal R}_{multi}^{(ini)}\right)^2,
\end{eqnarray}
which is consistent with the naive expectation given in
Eq.(\ref{bai-ve}).
See also Ref.\cite{End-multi-mat0} for an alternative approach.
A similar situation appears in strongly dissipating warm inflation, in
which the curvature 
perturbation is given by\footnote{In
ref.\cite{matsuda-warm} we considered warm inflation as an 
 interesting application of multi-field inflation and solved
the evolution of curvature perturbations about a FRW background 
metric, using the local conservation of energy-momentum. 
We found that there is a significant evolution of the curvature
perturbation if the curvature perturbation at horizon crossing 
is expressed by the standard gauge-invariant formula defined using the
total energy-momentum. The final result is consistent with the
$\delta N$ formula based on $\delta \phi$ and $\dot{\phi}$.}
\begin{eqnarray}
{\cal R}_{warm}^{(ini)}=-H\frac{\delta q}{\rho+P}=
H\frac{\dot{\phi}\delta \phi}
{\dot{\phi}^2+Ts}
\simeq H\frac{\dot{\phi} \delta \phi}
{Ts}.
\end{eqnarray}
Again, the evolution during inflation is crucial, since 
from the time-integral of a decaying component 
we find \cite{matsuda-warm},
\begin{eqnarray}
\left(\Delta{\cal R}_{warm}\right)^2 \simeq 
\left(\frac{r_\Gamma V_\phi}{(1+r_\Gamma)^2}\frac{\delta \phi}{
\dot{\phi}^2}\right)^2 \simeq H\frac{\delta \phi}{\dot{\phi}}\gg
\left({\cal R}_{warm}^{(ini)}\right)^2.
\end{eqnarray}
One might think that the result is trivial when the number of
e-foldings is determined exclusively by the inflaton.
However, just as in multi-field inflation where the significant
evolution of the curvature perturbation is important, in warm inflation 
the evolution of the curvature perturbation is very
important.\footnote{In ref.\cite{matsuda-warm} it is also found that a 
significant correction may arise from the 
perturbations of the radiation, which raises significant 
correction in the intermediate region ($r_\Gamma\simeq 1$).
For simplicity, the correction from the perturbations of the radiation
is not considered in this paper, although it is important in warm
inflation.} 

From the Langevin equation, the root-mean square fluctuation amplitude
of the inflaton field $\delta \phi$ after the freeze out 
is obtained to be\cite{warm-inflation-original}
\begin{eqnarray}
\label{pert-phi}
\delta \phi_{\Gamma >H}&\sim& 
(\Gamma H)^{1/4}T^{1/2}
\sim r_\Gamma^{1/4}r_T^{1/2}H\,\,\,\,\,\, (SDWI),\\
\delta \phi_{\Gamma<H}&\sim& 
(HT)^{1/2}
\sim r_T^{1/2}H\,\,\,\,\,\, (WDWI)
\end{eqnarray}
where $r_T$ denotes the ratio between $T$ and $H$, defined by
 $r_T\equiv T/H$, which is assumed to be $r_T >1$ during warm
 inflation.
The strength of the thermal damping is measured by the rate 
$r_\Gamma$ given by 
\begin{equation}
r_\Gamma\equiv \frac{\Gamma}{3H},
\end{equation}
which can be used to rewrite the field equation as
\begin{equation}
\ddot{\phi}+3H(1+r_\Gamma)\dot{\phi}+V(\phi,T)_\phi=0.
\end{equation}
For SDWI, the curvature perturbations created during warm inflation are
thus determined by the equation
\begin{eqnarray}
{\cal R}_{warm}\simeq H\frac{\delta \phi}{\dot{\phi}}
\simeq \frac{r_\Gamma^{3/4}}{r_T^{3/2}}.
\end{eqnarray}
For WDWI, it is given by 
\begin{eqnarray}
{\cal R}_{warm}
\simeq \frac{r_\Gamma^{1/2}}{r_T^{3/2}}.
\end{eqnarray}
Here we used $T^4 \sim r_\Gamma \dot{\phi}^2$, which is
explained in the next section.
An interesting property is expected from the above equations.
For remote inflation, one may naturally expect that many fields
are in thermal equilibrium and stay in their false vacuum states.
If there is a cascading symmetry breaking in the remote
sector as $T$ evolves during inflation, it raises a gap in the Hubble
parameter.
This may cause a transition from WDWI to SDWI, or simply a gap in the 
parameters $r_\Gamma$ and $r_T$.
Assuming instant decay in the remote sector, $T$ may have a
gap (sudden increase) at the transition.
This may happen if the energy released by the
transition ($\Delta V_{remote}$) raises the effective temperature $T$ 
for $\Delta V_{remote} > T^4 > H^4$.
Otherwise, one may assume that the change of the temperature $T$ is
negligible because of the slow-decay, and $\Gamma$ is a constant at the
transition. 
Even if both $T$ and $\Gamma$ are constant at the transition, a gap in
the curvature perturbation may appear from the gap in the Hubble
parameter. 
We investigate these interesting possibilities in this paper.
More significantly, the Hubble parameter may decrease with a
large ratio $H_{before}/H_{after}>1$ if the major part of the remote
sector decays.
Then, the increase of $r_\Gamma$ and $r_T$ caused by the
shift of the Hubble parameter leads to the amplification (or
suppression) of the scalar perturbation.

For example, we consider the simplest case in which the radiation released
from the remote sector is not significant due to the slow-decay, but the
shift of the Hubble parameter is not negligible.
For WDWI ($r_\Gamma <1$), the curvature perturbation is given by
\begin{equation}
{\cal R}_{warm}\propto r_\Gamma^{1/2}r_T^{-3/2}\sim H.
\end{equation}
For SDWI ($r_\Gamma >1$) it is given by
\begin{equation}
{\cal R}_{warm}\propto r_\Gamma^{3/4}r_T^{-3/2}\sim H^{3/4},
\end{equation}
These equations show that the curvature perturbation is suppressed
 at the gap. 
Note that a transition from WDWI to SDWI is also possible at the gap.
Contrary to scalar perturbations, tensor perturbations do not couple
strongly to the thermal background.
Gravitational waves are thus generated by quantum fluctuations as in
standard (supercooled) inflation\cite{warm-tensor}.
We thus find
\begin{equation}
A_g^2 \simeq \frac{H^2}{M_p^2},
\end{equation}
which decreases at the gap.
Here the ``gap'' is not always assumed to be large.
It may appear as a few percent of $T$ and $H$ to create a mild
scale-dependence. 

For the second example, we consider a gap characterized by
$\Delta T$ with the explicit form of the damping rate given by $\Gamma
\propto T^3/\phi^2$. 
If a small part of the false-vacuum energy is released by the phase
transition caused by a field in the multi-field remote sector,
 the gap in the Hubble parameter is
negligible but the gap in the temperature can be significant.
Since the damping rate depends explicitly on the temperature, 
the gap is accompanied by an increase in the damping rate.
Using $\Gamma \propto T^3$, there is a change in the curvature
perturbations 
\begin{eqnarray}
{\cal R}_{SDWI}&\propto& r_\Gamma^{3/4}r_T^{-3/2} \sim T^{3/4},\\
{\cal R}_{WDWI}&\propto& r_\Gamma^{1/2}r_T^{-3/2} \sim T^{0},
\end{eqnarray}
which may lead to a significant amplification of the curvature
perturbations at the corresponding cosmological scale.
The amplification, if it exists in the observable scale,
may appear as a temporal peak in the spectrum.
The tail of the peak is determined by the inflation attractor.
Finally, we note that the gap in the temperature caused by the phase
transition in the remote sector may {\bf appear even if the energy
density during inflation is not dominated by the remote sector.}
In this sense, the gap in the temperature is a common and very important
phenomenon in warm inflation, which requires further study.
Details of these topics are discussed in Sec.2.

\section{Remote inflation}

In conventional hybrid inflation, the typical form of the hybrid-type
potential is given by 
\begin{equation}
V(\phi,\sigma)=V(\phi) + \frac{g^2}{2} \sigma^2\phi^2
+\lambda\left(\sigma^2-M^2\right)^2.
\end{equation}
The vacuum energy during inflation is $V_0\simeq \lambda M^4$, which is 
defined by the false vacuum 
$\sigma=0$ and $\phi>\phi_c$.
Here the critical point is defined by
\begin{equation}
\phi_c\equiv \sqrt{\frac{\lambda}{g}}M.
\end{equation}
Because of the condition $\phi>\phi_c$, the motion of the inflaton field
must obey ``roll-in'' motion $\dot{\phi}<0$.
Another possibility for realizing hybrid inflation is to consider
``roll-out'' motion of the inflaton field considering a non-trivial
 form of the couplings.
The scenario that realizes ``roll-out'' motion is called the inverted
scenario of hybrid inflation\cite{inverted-hybrid}. 
The conditions related to the roll-in and roll-out motion are usually 
used to restrict inflaton potential.
Therefore, removing the conditions for hybrid inflation related to
``roll-in'' and ``roll-out'' is useful.\footnote{See
Ref.\cite{Clesse:2008pf} for recent arguments for the conditions for
successful hybrid inflation.} 

A natural construction of the hybrid-type potential in the string theory
is to consider brane distance as the inflaton.
The trigger field may be a tachyon that appears in the collision of 
a brane anti-brane pair\cite{brane-inflation-original}, or a field
localized on the target wall\cite{matsuda-non-tachonic}.
A thermal effect can cause an attractive force between branes, which can
be 
used to construct thermal brane inflation\cite{thermal-brane}, and to
construct models of brane inflation based on defect inflation
scenario\cite{matsuda-topological-brane} 
and brane inflation with roll-out motion \cite{matsuda-thermal-brane}.
Although brane inflation is not discussed explicitly in this paper, 
application of the remote inflation scenario to brane inflation is an
attractive idea.

Contrary to conventional hybrid inflation, either normal, inverted or
brane-motivated, remote inflation does not have direct couplings between
inflaton and trigger fields.
Namely, remote inflation is a kind of hybrid-like inflation but it does
not have the so-called hybrid-type potential.
The trigger fields $\sigma_i$ are placed in a remote sector and stays in
their false vacuum states during inflation, but they do not have direct
couplings with the inflaton field $\phi$. 
These trigger fields can stay in their false vacuum states because of 
the thermal effect from the background radiation. 
The background radiation is negligible in supercooled inflation, but it
may have a significant effect in warm ($T>H$) inflation.
A typical situation is shown in figure 1, in which either roll-in or
roll-out potential for inflaton are considered.

\begin{figure}[h]
 \begin{center}
\begin{picture}(200,300)(100,170)
\resizebox{15cm}{!}{\includegraphics{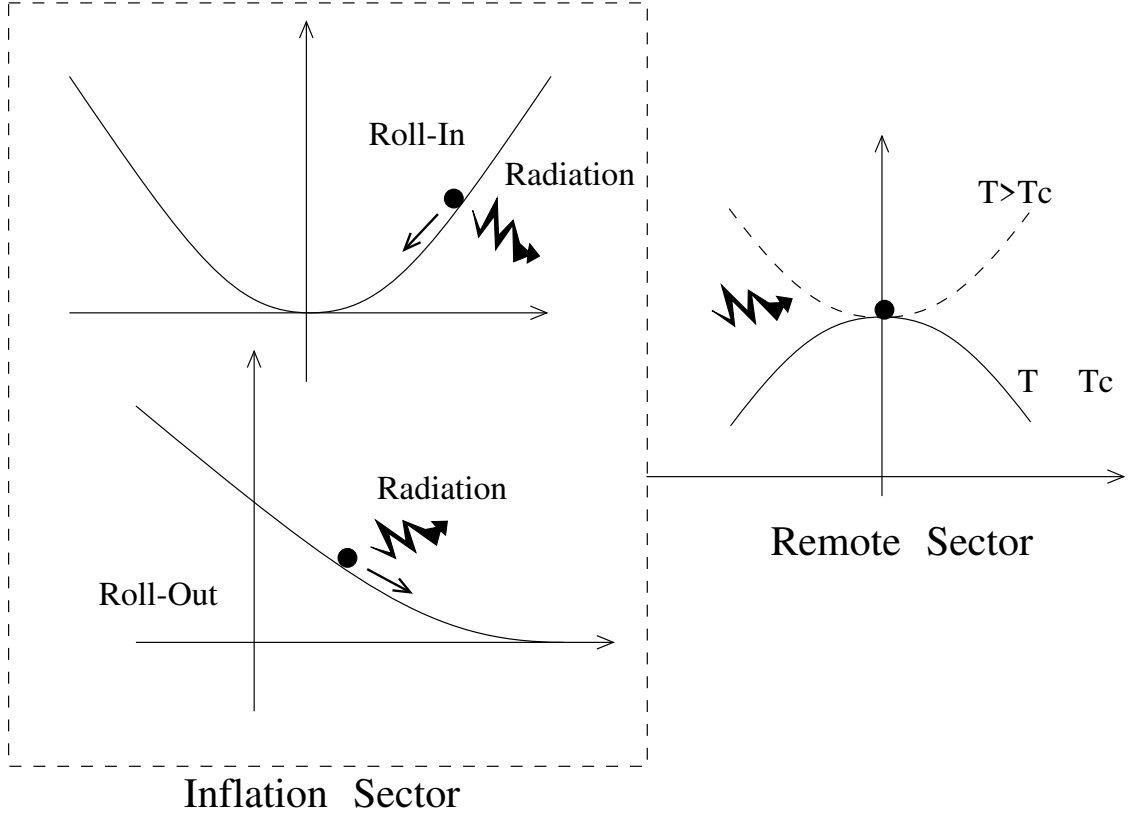}} 
\end{picture}
\label{fig:mass-dom}
 \caption{Thermal effect raised by a dissipating inflaton field causes
symmetry restoration in a remote sector, where the false-vacuum
potential energy dominates the energy density during inflation.
Remote inflation ends at a critical temperature when the
spontaneous symmetry breaking occurs in the remote sector.
Inflaton may either roll-in or roll-out on arbitrary inflaton potential.
Contrary to conventional hybrid scenario, direct coupling between
inflaton and a trigger field is not required for remote inflation.} 
 \end{center}
\end{figure}

In order to show how slow-roll inflation is realized in the warm 
inflation scenario, we first consider a homogeneous inflaton field
interacting with thermal radiation\cite{Hall-Moss-Berera}.
The thermodynamic potential is given by
\begin{equation}
V(\phi, T)= -\frac{\pi^2}{90}g_*T^4 + \frac{m(\phi,T)\phi^2}{2}
+V_0(\phi),
\end{equation}
where $g_*$ is the effective number of thermal particles.
The evolution equation for the inflaton field is given by
\begin{equation}
\ddot{\phi}+(3H+\Gamma)\dot{\phi}+V(\phi,T)_\phi=0,
\end{equation}
where $\Gamma$ is the damping term and $H$ is the expansion rate of the
Universe. 
Here the subscript denotes the derivative with respect to the field.
The strength of the thermal damping is measured by the rate $r_\Gamma$
 given by $r_\Gamma\equiv \frac{\Gamma}{3H}$,
which can be used to rewrite the field equation as
\begin{equation}
\ddot{\phi}+3H(1+r_\Gamma)\dot{\phi}+V(\phi,T)_\phi=0.
\end{equation}
The strongly dissipating phase of warm inflaton is defined by
$r_\Gamma\gg 1$. 
The production of entropy during inflation is caused by the dissipation
of the inflaton motion, where the entropy $s$ is defined by the
thermodynamic equation;
\begin{equation}
s\equiv - V(\phi,T)_T,
\end{equation}
where the subscript denotes the derivative with respect to the
temperature.
For our calculation, we need explicit forms of the energy density
$\rho$ and the pressure $p$, which are given by
\begin{eqnarray}
\rho &=& K + V + Ts\nonumber\\
p&=& K-V,
\end{eqnarray}
where $K\equiv\frac{1}{2}\dot{\phi}^2$ is the contribution from the 
kinetic term of the inflaton field, and
the Friedman equation follows
\begin{equation}
H^2 = \frac{1}{3M_p^2}\left(K+V+Ts\right).
\end{equation}
Here we assumed that inflaton has the standard kinetic term.
The slow-roll approximation is very useful in estimating the order of
magnitude of the physical quantities:
\begin{eqnarray}
\label{slou}
\dot{\phi}&\simeq& -\frac{V_\phi}{3H(1+r_\Gamma)}\\
\label{rel-warm}
Ts&\simeq& 2 r_\Gamma K\\
\rho+P &=& 2K +Ts \simeq 2(1+r_\Gamma)K,
\end{eqnarray}
where the second equation is obtained from the evolution equation for the
radiation energy density.
The slow-roll trajectory of inflation for $\phi$ and $T$ 
is determined by Eqs.(\ref{slou})
and (\ref{rel-warm}).
The gauge-invariant combinations for the curvature
perturbation are defined by
\begin{eqnarray}
\label{zeta-org}
\zeta &=&-\psi -H\frac{\delta \rho}{\dot{\rho}}\nonumber\\
{\cal R}&=& \psi -H\frac{\delta q}{\rho+p},
\end{eqnarray}
where $\delta q =-\dot{\phi}\delta
\phi$  is the momentum perturbation satisfying
\begin{equation}
\epsilon_m=\delta \rho-3H \delta q.
\end{equation}
Here $\epsilon_m$ is the perturbation of the comoving density,
satisfying the evolution equation 
\begin{equation}
\label{decay-comv}
\epsilon_m =-\frac{1}{4\pi G}\frac{k^2}{a^2}\Psi,
\end{equation}
where $\Psi$ is related to the shear perturbation and assumed to be
finite.
From the above definitions, $\delta q=0$ hypersurfaces are
identified with uniform density hypersurfaces ($\delta \rho=0$) 
at large scales due to the relation $\epsilon_m\equiv \delta
\rho-3H\delta q\simeq 0$, 
where these two definitions of the curvature perturbation are identified
($\zeta\sim -{\cal R}$). 
We considered linear scalar perturbations of a
Friedman-Robertson-Walker(FRW) background:
\begin{equation}
ds^2=-(1+2A)dt^2 + 2a^2(t)\nabla_i B dx^i dt +a^2(t)
[(1-2\psi)\gamma_{ij}+2\nabla_i\nabla_j E]dx^i dx^j.
\end{equation}
Using the above definitions based on the energy-momentum, we find that
 the comoving curvature perturbation created during warm inflaton
on spatially flat hypersurfaces $\psi=0$ is expressed as
\begin{eqnarray}
\label{normal-curv}
{\cal R}&=&  -H\frac{\delta q}{\rho+p}\nonumber\\
&=&  -H\frac{\delta q}{\dot{\phi}^2 + Ts}\nonumber\\
&\sim &  -H\frac{\delta q}{\dot{\phi}^2}\times (1+r_\Gamma)^{-1}.
\end{eqnarray}
The result is obviously much smaller than the result expected from the
perturbation of the number of e-foldings based on $\delta \phi$ and
$\dot{\phi}$, which is given by
\begin{equation}
\label{delta-multi}
\delta N\sim H\frac{\delta \phi}{\dot{\phi}}.
\end{equation}
As we discussed in Sec.1, a similar situation appears for multi-field
inflation, and the solution to this problem is given by the
evolution of the curvature perturbations.
Namely, for multi-field inflation a significant evolution at the
``bend'' in the trajectory leads to 
\begin{equation}
\Delta {\cal R}\simeq H\frac{\delta \phi}{\dot{\phi}},
\end{equation}
which recovers the expected form of the curvature perturbations
 based on $\delta \phi$ and
$\dot{\phi}$.
For warm inflation, a significant evolution appears from the
time-integral of a decaying
component in the evolution equation.
Again, it leads to the result consistent with the expectation based on
$\delta \phi$ and $\dot{\phi}$. 
In this paper, we thus consider that the curvature perturbations created 
during warm inflation are given by
\begin{eqnarray}
{\cal R}&\sim & H\frac{\delta \phi}{\dot{\phi}},
\end{eqnarray}
which is used without mentioning the origin.

An important source of density perturbations in warm inflation is
thermal fluctuations, which causes significant difference from
supercooled inflation.
Namely, a comoving mode of thermal fluctuations during warm
inflation is created by thermal noise, which can lead to significant 
field perturbations $\delta \phi \gg H$. 
The behavior of a scalar field interacting with radiation can be
studied by using the Langevin equation
\begin{equation}
-\nabla \phi(x,t)+\Gamma \dot{\phi}(x,t)+V_\phi=\xi(x,t),
\end{equation}
where $\xi$ denotes a source of stochastic noise.
From the Langevin equation, the root-mean square fluctuation amplitude
of the inflaton field $\delta \phi$ after the freeze is obtained.
It is given by a simplified form 
$\delta \phi\sim (\Gamma H)^{1/4}T^{1/2}
\sim r_\Gamma^{1/4}r_T^{1/2}H$ for the typical ($\Gamma >H$) warm
inflation, and $\delta \phi\sim (H T)^{1/2}
\sim r_T^{1/2}H$ for weakly dissipating ($\Gamma <H$) warm
inflation.
Using the relation given in Eq,(\ref{rel-warm}),
 the scalar curvature perturbation is expressed as
\begin{eqnarray}
{\cal R}_{SDWI}&\sim & \frac{r_\Gamma^{3/4}}{r_T^{3/2}},\\
{\cal R}_{WDWI}&\propto& \frac{r_\Gamma^{1/2}}{r_T^{3/2}}.
\end{eqnarray}
Cosmological perturbations created during remote inflation 
are calculated using these basic equations of warm inflation.

\subsection{Single-field remote sector}
Remote inflation requires symmetry restoration in the remote sector,
which is caused by the radiation dissipated from the inflaton.
To illustrate some typical features of finite temperature effects in the
remote sector, here we consider a real scalar field and a potential: 
\begin{eqnarray}
\label{ini-the}
{\cal L}&=&\frac{1}{2}\partial_{\mu}\sigma\partial^{\mu}\sigma-V(\sigma)\nonumber\\
V(\sigma)&=&V_0-\frac{1}{2}m_\sigma^2\sigma^2+\frac{1}{4}\lambda\sigma^4,
\end{eqnarray}
where $V_0$ is tuned so that the cosmological constant vanishes at the
true minimum. 
The phenomenon of high-temperature symmetry restoration
can be understood by the finite-temperature effective potential given by
\cite{EU-book-kolb}
\begin{equation}
V_T(\sigma_c)=V(\sigma_c)+\frac{T^4}{2\pi^2}\int^{\infty}_{0}
dx \ln \left[1-\exp\left(
-\sqrt{x^2+\frac{-m_\sigma^2+3\lambda\sigma_c^2}{T^2}}
\right)
\right],
\end{equation}
where $V(\sigma_c)$ is the one-loop potential for zero-temperature with
the classical field $\sigma_c$:
\begin{equation}
V(\sigma_c)=-\frac{1}{2}m_\sigma^2\sigma_c^2+\frac{1}{4}\lambda\sigma_c^4
+\frac{1}{64\pi^2}\left(-m_\sigma^2+3\lambda\sigma_c^2\right)^2
\ln \left(\frac{-m_\sigma^2+3\lambda\sigma_c^2}{\mu^2}\right),
\end{equation}
where $\mu$ is a renormalization mass scale.
At high temperatures, $V_T$ can be expanded near the symmetric point
($\sigma_c=0$) as
\begin{equation}
V_T\simeq V(\sigma_c)+\frac{1}{8}\lambda T^2 \sigma_c^2+ {\cal O}(T^4),
\end{equation}
which suggests that the temperature-corrected effective mass at 
$\sigma_c=0$ changes sign at the critical temperature given by
\begin{equation}
T_c \simeq \frac{2m_\sigma}{\lambda^{1/2}}.
\end{equation}
More generally, one may introduce a background thermal bath which couples
to $\sigma$.
Assuming that the couplings of $\sigma$ to the fields in the background
thermal bath are more significant than the self-coupling, 
we obtain a typical form of the potential with a thermal correction term,
which is given by
\begin{equation}
V=V_0+\left(g^2 T^2 -\frac{1}{2}m_\sigma^2\right)\sigma^2+ ...,
\end{equation}
where $g$ denotes the effective coupling of $\sigma$ to the fields in the
thermal bath. 
In this case, the critical temperature is given by
\begin{equation}
T_c \simeq \frac{m_\sigma}{2g}.
\end{equation}
In this paper, we assume that the couplings to the background fields
are significant and $T_c$ is always given by
$T_c \simeq \frac{m_\sigma}{2g}$. 
We define the critical temperature for the fields in the remote sector by
$T_c^{(i)}\equiv m_{\sigma_i}/2g_i$.

\underline{\bf Single field, quartic potential ($i=1$ and $n=4$)}

Since the false-vacuum energy in the remote sector determines the energy
scale of inflation, $V_0$ in Eq.(\ref{ini-the}) must be determined
explicitly.
The simplest form of the remote sector would be given by a single-field 
quartic potential;
\begin{eqnarray}
V&=&\frac{1}{4}\lambda\left(\sigma^2-M_*^2\right)\equiv V_0
-\frac{1}{2}m_\sigma^2\sigma^2+\frac{1}{4}\lambda \sigma^4\nonumber\\
&& V_0\equiv \frac{1}{4}\lambda M_*^4\nonumber\\
&& m_\sigma^2 \equiv \lambda M_*^2,
\end{eqnarray}
where the false-vacuum state is stabilized for 
$T>T_c\equiv \sqrt{\lambda} M_*/2g$.
For the single-field remote sector model, we find
\begin{equation}
H^2 \simeq \frac{V_0}{3M_p^2} = \frac{\lambda M_*^4}{12M_p^2}.
\end{equation}
Since the false-vacuum energy $V_0$ in the remote sector must dominate
the energy density during warm inflation, we have a condition 
$V_0 \gg T^4>T_c^4$, which leads to a simple equation
\begin{equation}
\frac{\lambda}{4g^4}\ll 1.
\end{equation}

It is also interesting to consider if a phase transition in the remote
sector occurs during a warm inflationary phase characterized by $T>H$.
The condition for the phase transition is given by $T>T_c>H$, which
leads to
\begin{equation}
 \frac{\sqrt{\lambda} M_*}{2g}
 >  \sqrt{\frac{\lambda M_*^4}{12M_p}}.
\end{equation}
Then, a condition $M_* < \sqrt{3}M_p/g$ is obtained.

For $T_c<H<T$, the phase transition may occur in supercooled
inflation.
If the radiation is not supplied concurrently from the inflaton, the
model becomes thermal inflation\cite{EU-book}.

\underline{\bf Single field, non-renormalizable potential
 ($i=1$ and $n>4$)}

More generally, we may consider an effective symmetry in the remote
sector, which forbids (or reduces) the quartic interaction.
In this case the potential is given by
\begin{eqnarray}
V&=& V_0 -\frac{1}{2}m_\sigma^2\sigma^2
+\frac{\lambda_n}{n} \frac{\sigma^n}{M_p^{n-4}},
\nonumber\\
&& <\sigma> =
\left(\frac{m_\sigma^2M_p^{n-4}}{\lambda_n }\right)^{1/(n-2)}\nonumber\\
&& V_0 \simeq  \left(\frac{m_\sigma^{n}M_p^{n-4}}{\lambda_n}
\right)^{\frac{2}{n-2}}.
\end{eqnarray}
The condition for the false-vacuum domination $V_0 \gg T^4>T_c^4$
leads to 
\begin{equation}
\frac{m_\sigma}{M_p}\ll 
g^2\left(\frac{g^4}{\lambda_n}\right)^{1/(2n-8)}.
\end{equation}

Again, from $T_c = m_\sigma/2g>H $, we obtain an obvious condition
for the phase transition;
\begin{equation}
\frac{m_\sigma}{M_p}< \left(\frac{\lambda_n}{g}\right)^{1/4}.
\end{equation}

\underline{\bf Inhomogeneous end of warm inflation}

In standard inflation, the end of inflation is determined by $\phi_c$,
where the slow-roll condition is violated at $\phi=\phi_c$.
The end-boundary of remote inflation is determined by either $\phi_c$
or $T_c$.
If $\phi$ reaches $\phi_c$ ahead of $T$, the situation is precisely the
same as the conventional warm inflation.
However, if $T$ reaches $T_c$ ahead of $\phi$, $T_c$ in the remote
sector  determines the end of inflation.
Also, if there is an inhomogeneous phase transition\cite{IH-pt} 
 in the remote sector, it
causes an inhomogeneous end of warm inflation.
Contrary to the usual scenario of generating curvature perturbations at
the end of inflation\cite{End-Modulated, End-multi,
 End-multi-mat}, here a model is considered in which 
the inhomogeneities are not caused by
 the inflaton-field perturbations at the end ($\delta \phi_e \ne 0$), 
but by the spatial
inhomogeneities of the critical temperature ($\delta T_c$)
 in the remote sector. 
Here the sources of the inhomogeneities are light fields (moduli) 
in the model, which determines the value of $g$ and $m_\sigma$.
From the definition of the critical temperature, we find
\begin{equation}
\frac{\delta T_c}{T_c}=\frac{\delta m_\sigma}{m_\sigma}-
\frac{\delta g}{g},
\end{equation}
which does not vanish if $m_\sigma$ or $g$ are determined by the vacuum
expectation values of light fields, which obtain cosmological
fluctuations at the horizon exit.
The effect can be used to add significant non-Gaussianity
to the spectrum, but to find the explicit form of the curvature
perturbations we must solve 
\begin{equation}
\delta N_{end} \sim H \frac{\delta T_c}{\dot{T}}
\end{equation}
at the end of warm inflation.
Since the calculation is highly model-dependent and requires numerical
study, we do not discuss details of this issue in this paper.
However, from a heuristic argument we can estimate the magnitude of the
correction by
\begin{equation}
\frac{\delta T_c}{\dot{T}}\sim 
\frac{T_c}{\dot{T}}\left[\frac{\delta T_c}{T_c}\right]
\sim \frac{\phi}{\dot{\phi}}\left[\frac{\delta T_c}{T_c}\right]
\sim \frac{\phi}{\delta \phi}\left[\frac{\delta \phi}{\dot{\phi}}\right]
\left[\frac{\delta T_c}{T_c}\right],
\end{equation}
which can be large.
Here we assumed that $T$ is given by a polynomial of $\phi$, assuming
that $r_\Gamma \sim C_\phi T^3/\phi^2$ and 
$T^4\sim r_\Gamma \dot{\phi}^2$ with  the inflaton potential
given by $\sim \phi^n$. 
The creation of the curvature perturbations caused by the inhomogeneous
phase transition in the remote sector is thus important for the
calculation of the curvature perturbations and its anomalies in the
spectrum. 

\subsection{Multi-field remote sector and cascading phase transition}

The essence of remote inflation is discussed in the previous section
considering the typical situation with single-field remote
sector. 
At the horizon exit, the single-field model discussed above creates the
same spectrum as the conventional warm inflation.
However, for the creation of the cosmological perturbations at the
inhomogeneous end of inflation, the inhomogeneous phase transition 
in the remote sector may leave a significant signature in the spectrum.
Therefore, the creation of the perturbations at the
end of inflation is important in distinguishing the model.

As we will show in this subsection, the significance of remote inflation
is more apparent in its multi-field 
extension, where the gaps caused by the phase transitions in the remote
sector may leave significant signatures in the spectrum.
In general, generation of gaps and peaks in the spectrum is
difficult in conventional inflation.
We thus believe that observations of these
signatures are important clues to remote inflation.
More precisely, in standard inflation, gap generation is caused by
non-trivial interactions that become significant at a critical point 
$\phi=\phi_A$.
Otherwise, the gap in the spectrum is explained by the gap
in the inflaton potential.
These mechanisms are highly model-dependent and require many
additional sources for the critical point.
Therefore, we believe that our idea, in which the gap is caused by the
critical temperature, is more attractive than the usual scenarios.

The effect from the phase transition in the multi-field 
remote sector can be categorized in the following ways.
\begin{enumerate}
\item $\Delta T/T >1$ and $H_{before}/H_{after} >1$, 
      where the major component in 
      the remote sector decays and supplies significant radiation.
\item $\Delta T/T <1$ and $H_{before}/H_{after}> 1$, where the major
      component in 
      the remote sector decays but due to the slow-decay it
      does not raise the temperature.
\item $\Delta T/T >1$ and $H_{before}/H_{after}<1$, where a minor
      component in 
      the remote sector decays and quickly thermalizes during inflation.
      Note that the vacuum-energy domination by the remote sector is not
      necessary for 
      the situation. (i.e., the situation is based on a small phase
      transition in a remote sector, which can appear in conventional
      warm inflation.) 
\end{enumerate}
The conditions for the situations and the expected corrections to the
curvature perturbations are discussed in the followings, using
an explicit form of $\Gamma$: $\Gamma \sim C_\phi
T^3/\phi^2$. 

\underline{1. $\Delta T/T >1$ and $H_{before}/H_{after} >1$}

In this case, the component in the remote sector that decays at the
phase transition has dominated the false-vacuum energy until
the phase transition.
We consider the situation in which the peak of the temperature
($T_{Max}$) caused by the phase transition exceeds the temperature
before the phase transition, but the decay is not so fast as
to dominate the energy density of the Universe.
Denoting the cosmological quantities before and after the phase
transition by the subscripts $-$ and $+$,
the condition is given by
$V_{0+}>T_{Max}^4>T_{-}^4$.
Because of the gap in the Hubble parameter, the ratios
$r_T\equiv T/H$ and $r_\Gamma\equiv \Gamma /3H$ increase after the gap.
Moreover, the supplement of the radiation from the remote
sector enhances $r_T \propto T$ and $\Gamma \propto T^3$
just after the gap.
Here we consider the damping rate $\Gamma$ that is defined by
\begin{equation}
\Gamma \simeq C_\phi \frac{T^3}{\phi^2},
\end{equation}
where $C_\phi \gg 1$ is the constant determined by the number of fields
and the channels for the decay\cite{warm-gamma}.
For simplicity, we introduce rates defined by 
\begin{eqnarray}
R_{T} &\equiv& \frac{T_{+}}{T_{-}}>1,\\
R_{H} &\equiv& \frac{H_{+}}{H_{-}}<1.
\end{eqnarray}
We thus find 
\begin{eqnarray}
\frac{r_{T+}}{r_{T-}}&=&\frac{R_{T}}{R_{H}},\\
\frac{r_{\Gamma +}}{r_{\Gamma -}}&=&\frac{R_{T}^3}{R_H},
\end{eqnarray}
which lead to the ratio of the curvature perturbations
\begin{eqnarray}
\frac{{\cal R}_+^{(WDWI)}}{{\cal R}_-^{(WDWI)}}&\simeq &
\frac{r_{\Gamma +}^{1/2} r_{T+}^{-3/2}}
{r_{\Gamma -}^{1/2} r_{T-}^{-3/2}}\simeq  R_{H},\\
\frac{{\cal R}_+^{(SDWI)}}{{\cal R}_-^{(SDWI)}}&\simeq &
\frac{r_{\Gamma +}^{3/4} r_{T+}^{-3/2}}
{r_{\Gamma -}^{3/4} r_{T-}^{-3/2}}
\simeq 
R_{H}^{3/4} R_{T}^{3/4},
\end{eqnarray}
which show that in WDWI the gap always causes suppression
of the curvature perturbation, while in SDWI amplification/suppression
 is determined by the model.
This typical characteristic feature can be used to distinguish SDWI from
WDWI. 
If the enhancement of $\Gamma$ is significant, the situation with 
$r_{\Gamma +} > 1 > r_{\Gamma -}$ is possible.
The instant transition from WDWI to SDWI leads to 
\begin{eqnarray}
\frac{{\cal R}_+^{(SDWI)}}{{\cal R}_-^{(WDWI)}}
&\simeq & \frac{r_{\Gamma +}^{3/4} r_{T+}^{-3/2}}
{r_{\Gamma -}^{1/2} r_{T-}^{-3/2}}\simeq
R_H  r_{\Gamma +}^{1/4}.
\end{eqnarray}
Again, the gap may cause either suppression or amplification.

\underline{2. $\Delta T/T <1$ and  $H_{before}/H_{after} >1$}

In this case, the component in the remote sector that decays at the
phase transition has dominated the false-vacuum energy,
but the creation of the radiation from the phase transition is very slow
 and thus the temperature is not raised significantly 
after the phase transition.
Therefore, we find $R_T\simeq 1$ and $R_H <1$ in this situation.
We thus find 
\begin{eqnarray}
\frac{r_{T+}}{r_{T-}}&=&\frac{1}{R_{H}},\\
\frac{r_{\Gamma +}}{r_{\Gamma -}}&=&\frac{1}{R_H},
\end{eqnarray}
which lead to the curvature perturbations
\begin{eqnarray}
\frac{{\cal R}_+^{(WDWI)}}{{\cal R}_-^{(WDWI)}}&\simeq &
 R_{H}<1,\\
\frac{{\cal R}_+^{(SDWI)}}{{\cal R}_-^{(SDWI)}}&\simeq &
R_{H}^{3/4}<1,
\end{eqnarray}
which show that in both cases the gap causes suppression
of the curvature perturbation.
If the enhancement of $\Gamma$ is significant, the situation with 
$r_{\Gamma +} > 1 > r_{\Gamma -}$ is possible.
The transition from WDWI to SDWI leads to 
\begin{eqnarray}
\frac{{\cal R}_+^{(SDWI)}}{{\cal R}_-^{(WDWI)}}
&\simeq & R_{H}r_{\Gamma +}^{1/4}.
\end{eqnarray}

\underline{3. $\Delta T/T >1$ and  $H_{before}/H_{after} <1$}

In this case, the component in the remote sector that decays at the
phase transition has not dominated the false-vacuum energy.
The same situation can be considered in standard warm inflation.
Here we assume for simplicity that the creation of the radiation from
the phase transition is rapid and the radiation is quickly raised after
the phase transition.
We consider the situation in which the peak of the temperature
($T_{Max}$) caused by the phase transition exceeds the temperature
before the phase transition.
Because there is no significant gap in the Hubble parameter, the ratio
$R_H$ is given by $R_H\simeq 1$ and can be neglected.
The creation of the radiation from the remote
sector enhances $r_T \propto T$ and $\Gamma \propto T^3$
just after the gap, which leads to $R_T>1$.
We thus find 
\begin{eqnarray}
\frac{r_{T+}}{r_{T-}}&=&R_{T},\\
\frac{r_{\Gamma +}}{r_{\Gamma -}}&=&R_{T}^3,
\end{eqnarray}
which lead to the curvature perturbations
\begin{eqnarray}
\frac{{\cal R}_+^{(WDWI)}}{{\cal R}_-^{(WDWI)}}&\simeq &
1,\\
\frac{{\cal R}_+^{(SDWI)}}{{\cal R}_-^{(SDWI)}}&\simeq &
 R_{T}^{3/4},
\end{eqnarray}
which show that for both WDWI and SDWI the gap always causes significant
amplification of the curvature perturbation.
If the enhancement of $\Gamma$ is significant, the situation with 
$r_{\Gamma +} > 1 > r_{\Gamma -}$ is possible.
The instant transition from WDWI to SDWI leads to 
\begin{eqnarray}
\frac{{\cal R}_+^{(SDWI)}}{{\cal R}_-^{(WDWI)}}
&\simeq & r_{\Gamma +}^{1/4}.
\end{eqnarray}

\subsection{Evolution of the non-equilibrium state}
From the evolution equation for the radiation energy density, the
radiation concurrently created from the inflaton is given by
\begin{equation}
\tilde{\rho}_R \sim r_\Gamma \dot{\phi}^2
\simeq  r_\Gamma \left(\frac{V_\phi}{3H(1+r_\Gamma)}\right)^2.
\end{equation}
Assuming that $r_\Gamma$ is given by the standard form 
$r_\Gamma \simeq C_\phi T^3/(3H\phi^2)$, the above equation gives
\begin{eqnarray}
\tilde{\rho}_R^{(WDWI)} &\propto& T^3\\
\tilde{\rho}_R^{(SDWI)} &\propto& T^{-3},
\end{eqnarray}
which show that the temporal increase in the radiation energy density
caused by the decay in the remote sector leads to
(1) enhancement of $\tilde{\rho}_R$ in WDWI or
(2) suppression of $\tilde{\rho}_R$ in SDWI.
As a result, the tail of the gap (the typical time-length of the
diffusion) in WDWI is typically longer than in SDWI.

Considering the attractor of the trajectory $(\phi,T)$ from the 
equations of warm inflation given in this section, finding
two independent solutions is possible, although it depends on the
details of the inflaton potential.
The transition between different trajectories is usually disregarded
because it 
is quite unlikely in conventional warm inflation.
However, in remote inflation the transition from a low-temperature
trajectory to a high-temperature trajectory may occur after the phase
transition in the remote sector.
The phase transition in the remote sector may also cause significant
change in the Hubble parameter that may also lead to the change of the
trajectory. 

\subsection{Cosmological defects}
Besides the complexities of the decaying process in the remote sector,
production of cosmological defects during inflation may leave 
significant signatures in the spectrum\cite{Jeannerot:2003qv}.
An interesting possibility that may arise in the context of an
inflationary
scenario\cite{vilenkin_book} is that monopoles are formed during
inflation but are not completely inflated away.
Then, at the confinement phase transition after inflation,
 strings connecting monopoles (or necklaces) are formed,
whose length is initially much longer than the Hubble radius.
Such strings may leave significant cosmological
signatures\cite{vilenkin_book, matsuda-PBH}, such as dark matter and
primordial black holes.
Besides strings and monopoles and their hybrids\cite{matsuda-necklaces},
cosmological domain walls may also appear when a discrete
symmetry is broken at the phase transition.
The creation of cosmological domain walls may cause serious cosmological
problem. 
Cosmological domain walls can decay safely before domination if
they are made unstable by a bias between the quasi-degenerated vacua,
which can be induced by an effective interaction term that breaks the
discrete symmetry.  
Note that for supergravity, domain walls caused by the R-symmetry are
naturally safe because the supergravity interaction causes the required
bias \cite{matsuda-dwall}. 
However, the instability caused by the explicit breaking term may not
work effectively to erase these domain walls before domination, because
the typical distance between domain walls created during
inflation is much larger than the Hubble radius. 
Namely, the distance between domain walls may still be larger than
the Hubble radius at the time when they must decay for successful
cosmology. 
This condition puts a trivial bound on the cosmological time of the
domain-wall creation during inflation. 
On the other hand, if such domain walls are generated sufficiently
late during the inflationary epoch and decay safely before domination,
they can be used to create significant amplification of the small-scale
density perturbations.
These topological defects (monopoles, strings, domain walls and
others\cite{sakai-TIT}) 
may put bounds on the phase transition, but at the same time they can
be used to create dark matter candidates, primordial black holes and 
small-scale density perturbations. 

\section{Conclusion and discussion}
In this paper we proposed a new scenario of hybrid-like inflation.
In remote inflation, radiation raised continuously by a dissipating
inflaton 
field causes symmetry restoration in a remote sector, and the
false-vacuum energy of the remote sector dominates the energy density 
 during inflation. 
The situation is similar to hybrid inflation, but the hybrid-type
potential is not required in this scenario.
Remote inflation is terminated when the temperature reaches the critical
 temperature, or when the slow-roll condition is violated.
Of course, the scenario is identical to thermal inflation if the
radiation is not supplied by the inflaton.
Without introducing additional couplings, the inflaton field may roll-out
like an inverted-hybrid model or quintessential inflation.
Significant signatures of remote inflation can be observed in the
 spectrum.
For a single-field remote sector, the difference is caused by the
inhomogeneous phase transition in the remote sector.
For multi-field extension of the remote sector, successive phase
transition is possible and it may cause significant difference in the
spectrum. 
Remote inflation can predict strong amplification or suppression
of small-scale perturbations, which may appear without introducing
multiple inflation.
Note that the phase transition in the remote sector is possible during
standard warm inflaton, irrespective of the domination by the
false-vacuum energy in the remote sector.

The inflaton field may have a run-away potential, which can be
identified with quintessential potential.
In this sense, the idea of remote inflation also provides a hybrid-type
extension of quintessential inflation.

For definiteness, we consider a concrete model of single field
inflation with quartic potential and calculate the
curvature perturbations and the spectral index of the model.
Note first that the spectrum of remote inflation is basically the same
as the warm inflation with the hybrid-type potential.
The potential of the model is given by
\begin{eqnarray}
V(\phi)&=&V_0 +\frac{1}{4}\lambda \phi^4,
\end{eqnarray}
where $V_0$ is the false-vacuum energy of 
the remote sector.
Here we consider a flat potential with $\lambda \ll 1$.
We consider the damping rate $\Gamma$;
\begin{equation}
\Gamma \simeq C_\phi \frac{T^3}{\phi^2}.
\end{equation}
For strongly dissipating phase, the slow-roll velocity is given by
\begin{equation}
\dot{\phi}\simeq \frac{V_\phi}{\Gamma} \simeq
\frac{\lambda \phi^5}{C_\phi T^3},
\end{equation}
which leads to the temperature during warm inflation
\begin{equation}
C_r T^7 \simeq 
\frac{\lambda^2 \phi^8}{4H C_\phi}.
\end{equation}
Here $C_r\equiv g_*\pi^2/30$ is the Stefan-Boltzmann constant.
A useful expression is
\begin{equation}
 \left(\frac{H}{\phi}\right)\simeq 
\frac{\lambda^2 }{4C_r C_\phi}\left(\frac{\phi}{T}\right)^{7}.
\end{equation}
From these equations, the curvature perturbation is obtained as
\begin{equation}
{\cal R}\simeq \frac{r_\Gamma^{3/4}}{r_T^{3/2}}
\simeq \left(\frac{C_\phi TH}{\phi^2}\right)^{3/4}
\sim \lambda^{3/2}C_r^{-3/4}\left(\frac{\phi}{T}\right)^{9/2}.
\end{equation}
Obviously, a tiny $\lambda$ is required for warm inflation.
Note that remote inflation works without the hybrid-type potential and
may solve 
problems in supercooled inflation, however it 
cannot be free of the conditions related to warm
inflation.
In ref.\cite{index-warm}, the spectral index of the curvature
perturbations in warm inflation has been studied for general form of the
dissipative coefficient.
The straightforward calculation of the spectral index using the useful
expression in Ref.\cite{index-warm} gives
\begin{equation}
n_s-1 \simeq -\frac{3\eta}{r}\left[-\frac{1}{7}+\frac{2\lambda}{7}
\frac{\phi^4}{V_0}\right]\sim\frac{3}{7}\frac{\eta}{r},
\end{equation}
where $r$ is given by
\begin{equation}
r\simeq \left(\frac{120}{\pi^2 g_*}\right)^{3/7} 
C_\phi^{4/7} \lambda ^{6/7}
\left(\frac{\phi}{H}\right)^{10/7}.
\end{equation}
The running of the spectral index for this model gives $dn_s/d\ln
k<0$\cite{index-warm}.\footnote{The spectral index $0.976<n_s<1.085$ is
allowed for the running of the spectral index $-0.065< dn_s/d\ln
k<-0.009$\cite{wmap5}.}

A more interesting and realistic application of remote inflation will be
discussed in Ref.\cite{sneutrino-remote}.

\section{Acknowledgment}
We wish to thank K.Shima for encouragement, and our colleagues at
Tokyo University for their kind hospitality.

\end{document}